\documentclass[reprint, twocolumn,notitlepage, superscriptaddress, nofootinbib]{revtex4-2}

\usepackage[english]{babel} 

\usepackage{microtype} 

\usepackage[svgnames]{xcolor} 

\usepackage{graphicx} 

\usepackage{enumitem} 
\setlist{noitemsep} 
\usepackage{lastpage}
\usepackage{fancyhdr}
\usepackage{lipsum}

\usepackage{geometry} 

\usepackage[T1]{fontenc} 
\usepackage[utf8]{inputenc} 

\usepackage{XCharter} 

\newcommand{\change}[1]{\textcolor{black}{#1}}

\usepackage{natbib}
\usepackage[nottoc]{tocbibind}
\addto\captionsenglish{}
\usepackage[autostyle=true]{csquotes} 
\usepackage{amsmath}
\usepackage{amsfonts}
\usepackage{amssymb}
\usepackage{textgreek}
\usepackage{cleveref}
\usepackage{braket}
\usepackage{bbold}
\usepackage{dsfont}
\usepackage{mathtools}

\usepackage{amsthm}
 
\theoremstyle{definition}

\usepackage{listings}
\usepackage{fancyhdr} 
\usepackage{lastpage}
\begin{document}
\title{Stress correlation function and linear response of Brownian particles}

\author{Florian Vogel}
\author{Matthias Fuchs} \affiliation{University of Konstanz - D-78457 Konstanz, Germany} 

\begin{abstract}
\textbf{Abstract-} We determine the nonlocal stress autocorrelation tensor in an homogeneous and isotropic system of interacting Brownian particles starting from the  Smoluchowski equation of the configurational probability density.  In order to relate stresses to particle displacements as appropriate in viscoelastic states, we go beyond the usual hydrodynamic description obtained in the Zwanzig-Mori projection-operator formalism by introducing the proper irreducible dynamics following  Cichocki and Hess, and Kawasaki.  Differently from these authors, we include transverse contributions as well. This recovers the expression for the stress autocorrelation including the elastic terms in solid states as found for Newtonian and Langevin systems, in case that those are evaluated in the overdamped limit. Finally, we argue that the found  memory function reduces to the shear and bulk viscosity in the hydrodynamic limit of smooth and slow fluctuations and derive the corresponding hydrodynamic equations. 
\end{abstract}

\date{\today} 

\begin{widetext}
\maketitle  
\end{widetext}

\section{Introduction}
Stress fluctuations play an important role in viscoelastic fluids, and understanding their spatio-temporal patterns remains an open question when starting from first principles \cite{Nicolas2018}.  
A system of interacting Brownian particles can be used to model the dynamics of concentrated colloidal dispersions \cite{Dhont}. While instantaneous solvent mediated interactions are neglected, the collective effects arising from steric particle interactions can be analyzed \cite{Marenne2017}. 
In the present work, the linear response of the local stress tensor $\pmb{\sigma}(\pmb{r},t)$ to an external velocity field $\pmb{v}^{ext}(\pmb{r'},t')$ at a distant space-time point is investigated in such a model of an overdamped colloidal  system. The main question is, whether precursors of the elastic properties of a colloidal glass already arise in the underlying fluid-like dynamics. The elastic response decays as $1/|\pmb{r}-\pmb{r'}|^3$ \cite{Eshelby1957,Nicolas2018}, while the fluid one is short-ranged.

This question was already considered in \cite{Vogel}. There, a set of  Langevin's equations of motion for the individual colloidal particles was investigated, which leads to a time evolution of the probability distribution function that is governed by the Klein-Kramers equation. It describes the dynamics in the phase space of the positions and velocities of the colloidal particles. When applying a Zwanzig-Mori projection formalism, it was argued that the coupling of the shear stress to the transverse current flow has to be taken into account, to  obtain the correct long-lived and long-ranged correlations in the supercooled state expected from the Newtonian case \cite{Maieretall2,Maieretall}. Only  based on this projection, the overdamped case and consequently the formation of colloidal solids could be considered. The long-ranged elastic stress pattern then develops via the strain correlations \cite{Illing2016,Hassani2018}, which enter as the current gradient is  the time-derivative of the strain.\\

In the present work, we take the alternative approach and start from an overdamped colloidal system governed by the Smoluchowski equation, which can be obtained from the overdamped limit of the Klein-Kramers equation \cite{HessKlein}. Here, the dynamics of the particles is described only in terms of their positions, as particle velocities have already relaxed into equilibrium. Thus, the question on defining the stress and consecutively viscosity via a momentum current arises \cite{Kruger2018}. Again, we first consider the hydrodynamic conserved variables within the Zwanzig-Mori formalism \cite{Zwanzig,Mori1965}.  Then, in order to relate the stress correlation to the particle displacement as previously found important, we generalize the work by Cichocki and Hess \cite{ChichockiHess} on the memory function for the dynamic structure factor by including transverse components as well.  Kawasaki \cite{Kawasaki} had already pointed out the curtailment to longitudinal terms.   Our ansatz naturally leads to the complete coupling of the tensorial stress fluctuations to the vectorial particle displacements. In the end, we regain the formally identical expression for the stress autocorrelation and therefore for the linear response of the stress tensor as in systems where velocities are kept as dynamical variables \cite{Vogel, Maieretall}. In the final chapter, we argue that the irreducible memory kernel indeed is the  correct one since it can be related to the \textit{Green-Kubo} transport coefficients, viz.~the shear and bulk viscosities. Since the generalized viscosity should be defined as the response of the local stress to the local current gradients \cite{Martin}, we have to reintroduce the flux as a dynamical variable. This is done via a linear response and a hydrodynamic approach. Both lead to consistent results. 

 In summary,  the coarse grained dynamics of dense colloidal dispersions can be obtained by properly identifying the irreducible Brownian dynamics; it yields the Brownian viscosity \cite{Foss2000} as transport coefficient.  Starting on the Brownian level appears more efficient than overdamping a calculation containing particle momenta.

\section{Brownian $N$-particle system}
We consider a  system containing \textit{N} interacting identical spherical particles performing random motion in a continuum background solvent. Usually, such a system is described with the locations $\{\pmb{r}_j\}^N$ and the momenta $\{\pmb{p}_j\}^N$ of the particles. In \cite{Vogel}, we used such a description to follow the system through the glass transition, and only after obtaining a general expression, the overdamped limit was considered. Here, we start with the overdamped simplification and consider $N$ interacting particles performing Brownian motion. Thus,  the state of the system can  be specified by the positions alone and is given by the \textit{N}-particle phase-space distribution \change{ $\rho(\Gamma,t)=\rho(\{\pmb{r}_j\}^N,t)$} at time $t$ for a fixed temperature $T$ and volume $V$. Thus, the Klein-Kramer's operator $\Omega$ governing the time-evolution $\partial_t \rho= \Omega \rho$ reduces \cite{HessKlein} to the Smoluchowski operator
\begin{align}
    \label{Smolochowski-eq}
    \Omega= D_0 \sum_{j=1}^N \frac{\partial}{\partial \pmb{r}_j} \cdot \bigg( \frac{\partial}{\partial \pmb{r}_j} - \beta \pmb{F}_j \bigg)\;.
\end{align}
Here $D_0 = \frac{k_B T}{\gamma_0}$ denotes the Stokes-Einstein-Sutherland diffusion coefficient, $\beta=1/(k_B T)$ represents the inverse temperature and $\pmb{F}_j$ is the total force acting on the $j^{th}$ particle.   On the other hand, the time evolution of a dynamical variable $\pmb{A}$ is governed by the adjoined Smoluchowski operator $\Omega^\dagger$ with $\partial_t \pmb{A}(t) = \Omega^\dagger \pmb{A}(t)$ and \cite{Risken}
\begin{align}
    \Omega^\dagger  =D_0 \sum_{j=1}^N  \bigg( \frac{\partial}{\partial \pmb{r}_j} + \beta \pmb{F}_j \bigg)\cdot \frac{\partial}{\partial \pmb{r}_j}.
\end{align}
The canonical average of a variable $ \pmb{A}$
\begin{align}\label{def_Average}
    \braket{ \pmb{A}(t)}= \int d \Gamma  \pmb{A}(t,\Gamma) \rho_{eq}(\Gamma)
\end{align}
defines a metric on the space of dynamic variables and can be generalised to an inner product, which can also be referred  to as correlation function
\begin{align}
\label{eq:Def_Correlationfunction}
\begin{split}
    C_{AB}(\pmb{r}- \pmb{r}',t-t'):&=\braket{ \pmb{A}(\pmb{r},t) \pmb{B}^*(\pmb{r}',t')}\\&= \int d \Gamma \delta \pmb{A}(\pmb{r})e^{\Omega (t-t')}  \delta \pmb{B}^*(\pmb{r}')\;  \rho_{eq}\;, 
    \end{split}
\end{align}
which is traditionally used to capture the dynamics of the system.
Here $^*$ represents the complex conjugate and \change{$\rho_{eq}  \propto  \exp[-\beta U(\{\pmb{r}_j\}^N)]$} is the canonical Gibbs-Boltzmann distribution. Only fluctuations away from equilibrium enter $\delta A \coloneqq A-\langle A\rangle$, yet we will  denote this in the following only in cases where the average  is non-vanishing. For simplicity, it is assumed that the potential \change{$U(\{\pmb{r}_j\}^N)$} can be decomposed into the sum of pair potentials which only depend on the distance between two particles. \change{This gives rise to central forces}. With this potential, one finds that the Hamiltonian \change{ $H(\{\pmb{r}_j\}^N ) = U(\{\pmb{r}_j\}^N )$} is invariant under rotation, translation and inversion. Thus, a fluid state of the system thus is homogeneous, isotropic, and achiral. As a consequence of that, the symmetry-related decompositions of  correlation functions found in \cite{Gotze,Maier,Maieretall} also hold for this system.   Since we consider equilibrium states, detailed balance holds as well. \change{ This gives  the operator identity \cite{Risken} $
    \Omega (\rho_{eq}...) = \rho_{eq}  \Omega^\dagger(...). $}\\

\section{Conservation law and dynamic variables}
In order to study the slow dynamics of the system resulting from the conserved hydrodynamic fields, a Zwanzig-Mori decomposition \change{ \cite{Zwanzig,Mori1965}} of the dynamical variable of interest, the stress tensor $\pmb{\sigma}$, will be considered. \change{Motivated by earlier works \cite{Maieretall,Vogel} and having the density as the only conserved dynamical variable, we choose the latter as the only distinguished variable. The Fourier-modes are given by} 
\begin{align}\label{eq:def_density}
    \varrho(\pmb{q})= \sum_{j=1}^N e^{-i\pmb{q}\cdot\pmb{r}_j}\;.
\end{align}
Due to the homogeneity of the system, the average of the density vanishes for $\pmb{q} \neq 0$. So, one finds $\braket{\varrho(\pmb{q})}=N \delta (\pmb{q})$.
The conservation law for the density gives the Laplacian of the stress tensor \change{
\begin{align}
    \label{eq:ConservationLaw}
    \begin{split}
    \Omega^\dagger \varrho(\pmb{q}) 
    &= -D_0 \beta q_\alpha q_\beta \sum_{j=1}^N \Big(k_B T \delta_{\alpha \beta} + i \frac{q_\alpha }{q^2}F_{\beta}^j \Big) e^{-i\pmb{q}\cdot \pmb{r}_j}  \\
    &= \vcentcolon- i D_0 \beta q_\beta f_\beta(\pmb{q})\\
    &= \vcentcolon - D_0 \beta q_\alpha q_\beta \sigma_{\alpha \beta}(\pmb{q})\;.
    \end{split}
\end{align}
With the force field $f_\beta(\pmb{q}) =-i q_\alpha \sigma_{\alpha \beta}(\pmb{q})$.}
Here and in the following, Greek letters refer to spatial directions and the  Latin indices $j, k$ to the $j^{th}$ or $k^{th}$ particle. Also Einstein's sum-convention was used and will be used from now on for Greek indices.
The former equation defines the stress tensor $\pmb{\sigma}$ up to addable $\nabla\cdot \pmb{\sigma}=0$ terms, which are neglected in the following. \change{Equation \eqref{eq:ConservationLaw} is fulfilled by}
\begin{align}\label{eq:Def_StressTensor}
\begin{split}
    \pmb{\sigma}( \pmb{q})&= \sum_{j=1}^N k_B T  e^{- i \pmb{q} \cdot \pmb{r}_j }\mathds{1} \\ &+ \sum_{k,j \neq k } 
    \pmb{r}_{jk}\pmb{ F}_{jk} 
    \frac{\sin( \frac{\pmb{q} \cdot \pmb{r}_{jk}}{2})}{ \pmb{q} \cdot \pmb{r}_{jk}}e^{-i \frac{\pmb{q}}{2} \cdot (\pmb{r}_j + \pmb{r}_k)}\;.
    \end{split}
\end{align}
Here $\mathds{1} $ represents the unity matrix and $ \pmb{r}_{jk}= \pmb{r}_{j}- \pmb{r}_{k} $.\change{The central force acting from the $k^{th} $ onto the  $j^{th} $ particle is denoted by  $\pmb{F}_{jk}=F_{jk}\frac{\pmb{r}_{jk}}{r_{jk}}$.}
Equation \eqref{eq:Def_StressTensor} is essentially the Irving-Kirkwood stress tensor where the canonical average over the momenta has been performed \cite{HansenMcDonald}.  With the same argument as above, the average $\braket{\pmb{\sigma}(\pmb{q})}$ vanishes for $\pmb{q} \neq 0$. On the other hand,  one finds for $\pmb{q}=0$
\begin{align}
    \langle \sigma_{\alpha \beta}(\pmb{q}=0) \rangle =   N  k_B T \delta_{\alpha \beta} +  \braket{ \sum_{k } r^k_\alpha F^k_\beta(\Gamma)}.
\end{align}
Taking the trace of the right side relates the stress tensor to the virial expression for the pressure $p$ \cite{HansenMcDonald}
\begin{align}
  p \coloneqq\frac{1}{3V}\mathrm{Tr}[\pmb{\sigma}] = n k_B T + \frac{1}{3 V }\braket{ \sum_{k} \pmb{r}_{k} \cdot \pmb{F}_k},
\end{align}
with the number density $n\coloneqq\frac{N}{V}$.

\section{The Stress autocorrelation}
The main object of interest is the fourth-rank tensor of the stress autocorrelation
\begin{align}
    C_{\alpha \beta \gamma \delta}(\pmb{q},t): = \frac{\beta}{ V } \braket{ \sigma_{\alpha \beta} (\pmb{q})  e^{\Omega t} \sigma_{\gamma \delta}^* (\pmb{q})}\;.
\end{align}
Note, that the time independence of the Smoluchowski operator and the resulting invariance of the equilibrium distribution under time translation  allows  setting $t'=0$ in equation~\eqref{eq:Def_Correlationfunction}. The assumed homogeneity  causes the double Fourier-transformation $C(\pmb{q},\pmb{q}',t)=\mathcal{F}[C_{A(\pmb{r})B(\pmb{r}')}](\pmb{q},\pmb{q}')$ to be zero, unless $\pmb{q}=\pmb{q}'$ holds. \change{We use the convention $\mathcal{F}[f(\pmb{r})](\pmb{q})=\int_V d \pmb{r} f(\pmb{r})e^{-i \pmb{q} \cdot \pmb{r}}$. Since we consider finite but large systems, we take $\frac{(2 \pi)^3}{V} \to 0$ in the end, giving rise to continuous Fourier-modes}.\\

As can be seen in equation \eqref{eq:Def_StressTensor},  the stress tensor is symmetric, meaning $\sigma_{\alpha \beta}=\sigma_{\beta \alpha}$. \change{(While this holds straightforwardly for central forces, which we consider,  it was shown in Ref.~\cite{Martin1972} that  a symmetric stress tensor can be introduced also in the case of non-central forces.)} This gives rise to symmetry properties of the stress autocorrelation:\change{
\begin{align}
    \label{eq:Symmetry}
     C_{\alpha  \beta  \gamma \delta} (\pmb{q},t) =  C_{  \beta \alpha  \gamma \delta } (\pmb{q},t) = C_{ \gamma \delta  \alpha  \beta }( -\pmb{q},t)=C_{ \gamma \delta \alpha  \beta }( \pmb{q},t).
\end{align}}
Here, the first identity holds because of the symmetry of $\boldsymbol{\sigma}$, the second because of detailed balance, and the last identity holds due to the achirality of the system. 
Based on  the isotropy of the system,  the stress autocorrelation can be decomposed into five \change{functions} depending on the wavenumber $q$ only,  as it is also possible in molecular systems \cite{Lemaitre2015,Maieretall}. These functions generalize the two viscosities (equivalently Lam\'e coefficients) to finite frequencies and wavevectors. \\

For simplicity, the following analysis of the stress-tensor will be done in the Laplace-domain with the convention $f(z)=\int_0^\infty dt f(t)e^{-zt}$, which gives for the stress autocorrelation function 
\begin{align}
    C_{\alpha  \beta \gamma \delta } (\pmb{q},z)= \frac{\beta}{ V } \braket{ \sigma_{\alpha \beta} (\pmb{q})  R(z) \sigma_{\gamma \delta}^* (\pmb{q})}\;,
\end{align}
with the resolvent $R(z)=[z-\Omega]^{-1}$.  In the next section, the expression for $C_{\alpha  \beta \gamma \delta} (\pmb{q},z)$ within the Zwanzig-Mori formalism will be derived using \change{G\"otze's} decomposition \cite{Gotze} for the resolvent. 
\begin{align}
\label{eq:Def_ResolventDecomposition}
    R(z) = R'(z) + [1+R'(z) \Omega]P R(z) P [1+ \Omega R'(z)]\;.
\end{align}
Here, $P$ projects  on the \change{subspace of distinguished  variables} and $R'(z)=Q[z-Q\Omega Q]^{-1}Q$ denotes the reduced dynamics.  $Q=1-P$ projects on the subspace of \change{remaining} variables.
\subsection{Generalized Memory function}
Hydrodynamic conservation laws define the slow variables which need to be specified in a local equilibrium state  \cite{HansenMcDonald,HessKlein}. Since the density is the only conserved dynamic variable in a Brownian system, the subspace of slow variables is one dimensional. The associated projection operator reads
\begin{align}
    \label{eq:Def_ProjectionOperator}
    P= 1- Q = \frac{1}{N S_q} \ket{\varrho^*(\pmb{q})}\bra{\varrho(\pmb{q})}\;.
\end{align}
Here $S_q\coloneqq\frac{1}{N}\braket{\varrho(\pmb{q})\varrho^*(\pmb{q})}$ is the static structure factor. Using this projection and equation \eqref{eq:Def_ResolventDecomposition}, we will describe the considered system as two coupled subsystems. The first one deals with the slow or conserved  density, while the latter is concerned with the remaining fast variables having the density as a constant parameter. Generally in such Zwanzig-Mori decomposition, the subsystems are coupled via  memory functions $M_{mk}\propto \braket{A_m\Omega  R'(z)  \Omega A_k}$, where $A_{m/k}$ are \change{distinguished} variables.
\subsubsection{Dynamic structure factor} 
The dynamics of the one dimensional subspace of slow variables is governed by the density autocorrelation  \change{ $S(q,z)\coloneqq\frac{1}{N}\braket{\varrho(\pmb{q})R(z)\varrho^*(\pmb{q})}$ later referred to as dynamic structure factor}. Using the Zwanzig-Mori equation of motion \cite{HansenMcDonald}, one finds a first expression for the dynamic structure factor \cite{Ackerson1976}
\begin{align}
    \label{wrong_Strukturfaktor}
    S(q,z)= \frac{S_q}{z+\frac{D_0 q^2}{S_q}(1- \frac{D_0  \beta}{n q^2} q_\alpha q_\beta \widetilde{M}_{\alpha \beta \gamma \delta}(\pmb{q},z)\  q_\gamma q_\delta)},
\end{align}
\change{with} an explicit expression for a first memory kernel \change{
\begin{align}
    \label{eq:FirstMemory}
    \widetilde{M}_{\alpha  \beta  \gamma \delta} (\pmb{q},z)= \frac{\beta}{ V } \braket{ \sigma_{\alpha \beta} (\pmb{q})  R'(z) \sigma_{\gamma \delta}^* (\pmb{q})}\;.
\end{align}}
Equation \eqref{wrong_Strukturfaktor} suggests to introduce the longitudinal and \change{(for later reference)} transverse component of the memory function   $\widetilde{M}^\parallel$ and $\widetilde{M}^\perp$
\begin{align}
\label{eq:Def_Longitudianl_Transverse}
\begin{split}
\widetilde{M}^\parallel(q,z)&\coloneqq \frac{\beta}{V}\braket{\sigma^{\parallel}(\pmb{q})R'(z)\sigma^{\parallel*}(\pmb{q})} \\
\widetilde{M}^\perp(q,z)& \coloneqq\frac{\beta}{V}\braket{\sigma^{\perp}(\pmb{q})R'(z)\sigma^{\perp*}(\pmb{q})}. 
\end{split}
\end{align}
With the longitudinal- $\sigma^\parallel\coloneqq\frac{1}{q^2} \pmb{q} \cdot (\pmb{\sigma} \cdot \pmb{q})$ and the transverse component  $\sigma^\perp\coloneqq\frac{1}{q^3}  \pmb{a} \cdot \left( q^2  (\pmb{\sigma}\cdot\pmb{q})- \pmb{q}((\pmb{\sigma}\cdot\pmb{q})\cdot\pmb{q})\right)$ of the stress tensor. Here, $\pmb{a}$ is an arbitrary normalised vector, perpendicular to $\pmb{q}$, meaning $| \pmb{a}|=1$ and $\pmb{a} \cdot \pmb{q}=0$.   The longitudinal memory function appears in the dynamic structure factor. Note that correlation functions of two scalar quantities depend  only on the absolute value of the wavevector $q=|\pmb{q}|$, due to rotational invariance. \change{ The cross product of the parallel and perpendicular components vanishes due to the achirality of the system } \cite{Gotze}.

\subsubsection{Irreducible Memory function}

In \cite{ChichockiHess}, Cichocki and Hess argued that 
$\widetilde{M}^\parallel$ is not the correct memory function, since it can not be identified with the longitudinal viscosity $\eta^\parallel$ \cite{HessKlein} and since it is not \textit{irreducible} as intended for a memory-function. Even though, the structure of $R'$ is such, that the governed dynamics is orthogonal to $\varrho(\pmb{q})$, the second order spatial derivatives describing diffusive processes in \eqref{Smolochowski-eq} cause a non-trivial influence of the density evolution on the fast variables. \change{ 
Cichocki and Hess were able to show that a further projection is possible and an additional \textit{one-particle} reduction can be done by inserting the projector $Q_j = 1-P_j$, with $P_j = \ket{e^{i \pmb{q}\cdot \pmb{r}_j}}\bra{e^{-i \pmb{q} \cdot \pmb{r}_j}} $, in $\Omega$. This was  supported by Kawasaki \cite{Kawasaki} within a more general approach. He showed that generally in dissipative systems with detailed balance a further projection can be performed. Considering Brownian motion as an example, Kawasaki found the same results as in \cite{ChichockiHess}.}\\

\change{
Following these authors, we define the irreducible Smoluchowski operator
    \begin{align}
        \label{IrSO}
        \Omega^{irr} \coloneqq QD_0 \sum_{j=1}^N \partial_{\pmb{r}_j} Q_j \cdot \bigg( \partial_{\pmb{r}_j} - \beta \pmb{F}_j \bigg) Q \;.
    \end{align}
By exploiting that the particles are statistically equivalent, one can relate the irreducible operator to the reduced Smoluchowski operator
    \begin{align}
\label{eq:Smolochowski_Operator_Irreducible}
\begin{split}
    Q\Omega Q&= \Omega^{irr} - \frac{D_0 \beta^2}{N} q_\alpha q_\eta \ket{ Q\sigma_{\alpha \beta}^*(\pmb{q})} \bra{\sigma_{\eta \beta}(\pmb{q})Q}\\&= \Omega^{irr}- \frac{D_0 \beta^2}{N}\ket{Qf_\alpha^* (\pmb{q})}\bra{f_\alpha(\pmb{q})Q}\;.
    \end{split}
\end{align}
This can be done, since an expression as the one above will later only appear in averages over all particles.}
\change{The irreducible operator in \eqref{eq:Smolochowski_Operator_Irreducible} decouples the time evolution from the slow dynamics of the longitudinal and transverse forces, but,  as Kawasaki noted \cite{ChichockiHess,Kawasaki},  Cichocki and Hess only accounted for the longitudinal forces in the continuation of their work.  They neglected the transverse components in \eqref{eq:Smolochowski_Operator_Irreducible} and  assumed that $\Omega \approx \Omega^{irr} - \frac{D_0 \beta^2}{N} q^2\ket{ \sigma^{\parallel*}(\pmb{q})} \bra{\sigma^{\parallel}(\pmb{q})}$ or equivalently,  $\Omega \approx \Omega^{irr} - \frac{D_0 \beta^2}{N} \ket{ \pmb{f}^{\parallel*}(\pmb{q})} \cdot \bra{\pmb{f}^{\parallel}(\pmb{q})}$ holds; here $ \pmb{f}^{\parallel}(\pmb{q}) = \pmb{q}\; (\pmb{q} \cdot  \pmb{f}^{\parallel}(\pmb{q}) ) /q^2 $. Looking at \eqref{eq:Smolochowski_Operator_Irreducible}, this choice seems rather unnatural since the transverse components, viz.~ $\sigma^\perp(\pmb{q})$ or equivalently $\pmb{f}^\perp(\pmb{q})=\pmb{q} \times (\pmb{f}(\pmb{q}) \times \pmb{q})/q^2$, are simply disregarded. We proceed differently than \cite{ChichockiHess,Kawasaki}, by arguing that 
all of the stress components including the transverse ones have to be considered in equation \eqref{eq:Smolochowski_Operator_Irreducible}. 
    Our claim rests on the observation that the restriction to the longitudinal contributions was artificial and the full tensorial structure of the stress arises naturally, also e.g.~in confined fluids \cite{Schrack2020}. Gratifyingly, our generalization leads to the same expression for the stress autocorrelation as in overdamped Newtonian and Langevin systems \cite{Vogel, Maieretall, Maieretall2}. Furthermore, we will show in chapter \ref{sey:Response_function} that the memory function can  be identified with the viscosity.} \\
    
    \change{Equation \eqref{eq:Smolochowski_Operator_Irreducible}  leads to a relation between the reduced and the  irreducible resolvent which differs from the one in Refs.~\cite{ChichockiHess,Kawasaki}}.

\begin{align}
\label{eq:_Relation_Resolvents}
    R'(z)= R^{irr}(z) \Big(1-   \frac{D_0 \beta^2}{N} q_\alpha q_\eta \ket{ \sigma^*_{\alpha \beta}} \bra{\sigma_{\eta \beta} } R'(z)   \Big)\;,
\end{align}
with the irreducible resolvent equals $R^{irr}(z)\coloneqq Q [z -  \Omega^{irr} ]^{-1}Q$.
So  $\widetilde{M}^\parallel$ and $\widetilde{M}^\perp$ can be expressed in terms  of a second set of memory functions $M^\parallel $ and $M^\perp$ defined in analogy to equation~\eqref{eq:Def_Longitudianl_Transverse}. One gets
\begin{align}
\label{eq_Longi_Trans_Memoryfunction}
\begin{split}
 \widetilde{M}^\parallel(q,z) &=    \frac{M^\parallel(q,z)}{1+\frac{D_0 \beta q^2 }{n}M^\parallel(q,z)} \\
\widetilde{M}^\perp(q,z) & = \frac{M^\perp(q,z)}{1+\frac{D_0 \beta q^2}{n}M^\perp(q,z)}\;,
\end{split}
\end{align}
with the irreducible memory kernel that reads explicitly:
 \begin{align}
 \label{eq:Memoryfunction}
        M_{\alpha \beta \gamma \delta}(\pmb{q},z)\coloneqq \frac{\beta}{V} \braket{\sigma_{\alpha \beta}(\pmb{q}) R^{irr}(z)  \sigma_{\gamma \delta}^*(\pmb{q})} \;.
    \end{align}
    Note, that the symmetry relations \eqref{eq:Symmetry}    hold for the memory function as well.
With the upper line of equation \eqref{eq_Longi_Trans_Memoryfunction}, the formally identical expression for the dynamic structure factor from Ref.~\cite{HessKlein} can be obtained \change{
\begin{align}
\begin{split}
\label{eq:Strukturfactor}
    S(q,z) &= \frac{1}{N}\braket{\varrho(\pmb{q})R(z)\varrho^*(\pmb{q})}\\&= \frac{S_q}{z + \frac{D_0 q^2}{S_q}(1+q^2\frac{D_0 \beta }{n}M^\parallel(q,z))^{-1}}\;.
    \end{split}
\end{align}
}
\change{In the hydrodynamic limit, $q\to0$ and $z\to 0$ such that $z/q^2=$const., it describes the collective particle diffusion with the osmotic diffusion coefficient $D=D_0/S_0$. In viscoelastic fluids, the memory kernel encoding  a frequency and wavenumber dependent friction cannot be neglected and approximations are required to find $M^\parallel(q,z)$ \cite{Gotze}. }

\subsection{Projection operator decomposition of the stress autocorrelation}
In this section, an expression for the stress autocorrelation shall be derived, including certain terms which can then be interpreted as a coupling to the longitudinal and transverse displacement of the particles. Using  the resolvent identity \eqref{eq:Def_ResolventDecomposition}, one finds a first expression for the stress autocorrelation:
\begin{widetext}
\begin{align}
\label{unfertigeautocorrelation}
\begin{split}
    C_{\alpha \beta \gamma \delta}(\pmb{q},z)&=\widetilde{M}_{\alpha \beta \gamma \delta}(\pmb{q},z) + \frac{S(q,z)}{S_q^2}\frac{n}{\beta} \delta_{\alpha \beta} \delta_{\gamma \delta} - \frac{S(q,z)}{S_q^2}D_0 q_\eta q_\xi \Big[\delta_{\alpha \beta}\widetilde{M}_{\eta \xi \gamma \delta}(\pmb{q},z) + \widetilde{M}_{\alpha \beta \eta \xi}(\pmb{q},z)\delta_{\gamma \delta}\Big] \\&+ \frac{S(q,z)}{S_q^2}\frac{D_0^2 \beta}{n} q_\eta q_\xi q_\lambda q_\tau \widetilde{M}_{\alpha \beta \eta \xi}(\pmb{q},z) \widetilde{M}_{\lambda\tau \gamma \delta}(\pmb{q},z)\;.
    \end{split}
\end{align}
\end{widetext}
In order to get the stress autocorrelation appropriate for a Maxwellian viscoelastic fluid \cite{Maieretall2}, one has to express the reducible memory function $\widetilde{\pmb{M}}$ in terms of the irreducible one. Expressing the reduced dynamics with \eqref{eq:_Relation_Resolvents} twice gives
\begin{align}
\label{Decomposition1}
\begin{split}
        \widetilde{M}_{\alpha \beta \delta \gamma}&= M_{\alpha \beta \delta \gamma}\\&- \frac{D_0 \beta}{n}M_{\alpha \beta \eta \vartheta}q_\eta q_\lambda   M_{\zeta \lambda \gamma \delta}\bigg( \delta_{\zeta \vartheta} - \frac{D_0 \beta}{n}\widetilde{M}_{\vartheta \mu \xi \zeta } q_\xi q_\mu\bigg)\;.
\end{split}
\end{align}
Where a commutation relation of $\widetilde{\pmb{M}}\pmb{q}\pmb{q}\pmb{M}$ was used following from the operator identity
\begin{align}
\label{eq:OperatorIdentity}
\begin{split}
    [\pmb{A}+\pmb{B}]^{-1}&=\pmb{A}^{-1}\Big(1- \pmb{B}[\pmb{A}+\pmb{B}]^{-1} \Big)\\&=\Big(1- [\pmb{A}+\pmb{B}]^{-1}\pmb{B}\Big)\pmb{A}^{-1}.
    \end{split}
\end{align}
Note that the arguments will be suppressed from this point on, if they reduce the clarity.
The remaining reducible memory function in the bracket of equation \eqref{Decomposition1} can be decomposed into the longitudinal and transverse memory function using equations \eqref{eq:Def_Longitudianl_Transverse} and \eqref{eq_Longi_Trans_Memoryfunction}. This gives
\begin{align}
\begin{split}
    \label{eq:Decomposition_of_reducible_Memory}
     &\widetilde{M}_{\alpha \beta \delta \gamma}= M_{\alpha \beta \delta \gamma}- \frac{D_0 \beta}{n}M_{\alpha \beta \eta \vartheta}q_\eta q_\lambda M_{\zeta \lambda \gamma \delta}\\\times& \bigg( \frac{q_\zeta q_\vartheta}{q^2}\frac{1}{1+ \frac{D_0 \beta}{n}q^2 M^\parallel}+ \Big(\delta_{\zeta \vartheta}-\frac{q_\zeta q_\vartheta}{q^2} \Big) \frac{1}{1+ \frac{D_0 \beta}{n}q^2 M^\perp} \bigg).
     \end{split}
\end{align}
By multiplying with $\frac{S(q,z)}{S_q} \pmb{q}^T \pmb{q}^T$ from the left, exploiting that longitudinal and transverse components do not couple due to the \change{spatial symmetries}, and by inserting \eqref{eq:Strukturfactor}, one obtains
\begin{align}
    \label{eq:Second_used_decoposition_memory}
    \frac{S(q,z)}{S_q} q_\alpha q_\beta \widetilde{M}_{\alpha \beta \gamma \delta}= \frac{1}{z  + \frac{k_B T}{S_q \gamma_0}+ \frac{q^2 z}{n \gamma_0}M^\parallel}q_\alpha q_\beta M_{\alpha \beta \gamma \delta}\change{\;,}
\end{align}
and an analogous expression for $q_\gamma  q_\delta \widetilde{M}_{\alpha \beta \gamma \delta}$. To make the obtained relations more handy,
one can define two scalars:
\begin{align}
    \label{Def_Longitudinal_Transverse_Current}
    \begin{split}
          K^\parallel(q,z)&\coloneqq \frac{k_B T}{\gamma_0 + \frac{k_B T}{z S_q}q^2+\frac{q^2}{n}M^\parallel(q,z)} \;,  \\
           K^\perp(q,z)&\coloneqq \frac{k_B T}{\gamma_0+\frac{q^2}{n}M^\perp(q,z)}\;.
    \end{split}
\end{align}
Those quantities can be arranged in a  matrix 
\begin{align}
    K_{\alpha \beta} = \frac{q_\alpha q_\beta}{q^2}K^\parallel + \Big(\delta_{\alpha \beta}- \frac{q_\alpha q_\beta}{q^2}\Big) K^\perp.
\end{align}
In this way, one notices that $K^\parallel$ and $K^\perp$ can be obtained from the expressions for the parallel and transverse current autocorrelations, respectively, in a Langevin system \cite{Vogel,HessKlein} by neglecting the inertia term. Thus, the matrix $K_{\alpha\beta}$ agrees with the current autocorrelations of a Langevin (or Newtonian) system in  the overdamped approximation of the full dynamics. \change{In order to interpret $K_{\alpha\beta}$  physically, it can be connected to either  displacement correlations \cite{Illing2016} or to a force correlation matrix based on equation \eqref{eq:ConservationLaw}. Explicitly, using the equations \eqref{unfertigeautocorrelation} and  \eqref{eq:Decomposition_of_reducible_Memory}, one finds 
\begin{align}
    K_{\alpha \beta}= - \frac{1}{N \gamma_0^2}\braket{f_\alpha R(z) f_\beta } + D_0 \delta_{\alpha \beta}\; .
\end{align} Yet, keeping the current correlations in the following equations is remindful that stresses lead to particle motions which affect the stresses in turn. Moreover, it leads to the most concise equations.}\\


Inserting \eqref{eq:Decomposition_of_reducible_Memory}, \eqref{eq:Second_used_decoposition_memory} in \eqref{unfertigeautocorrelation} gives the expression for the stress autocorrelation:
\begin{widetext}
\begin{align}
\label{mainResult}
    \begin{split}
          C_{\alpha \beta \gamma \delta}(\pmb{q},z)= &M_{\alpha \beta \gamma \delta} +  \frac{S(q,z)}{S_q^2}\frac{n}{\beta} \delta_{\alpha \beta} \delta_{\gamma \delta} - \frac{1}{S_qz} \bigg[\delta_{\alpha \beta} M_{\eta \xi \gamma \delta}q_\xi q_\eta K^\parallel +M_{\alpha \beta \eta \xi }q_\xi  q_\eta K^\parallel \delta_{\gamma \delta} \bigg] 
     - \frac{\beta}{n} M_{\alpha \beta \eta \vartheta}q_\eta K_ { \vartheta \zeta}\, q_\lambda M_{\zeta \lambda \gamma \delta}.
    \end{split}
\end{align}
\end{widetext}
This decomposition of the stress autocorrelation naturally includes an \change{exact} contribution arising from the coupling of the stress to the conserved variables, viz.~the conserved density in the present case which enters via the dynamic structure factor $S(q,z)$. \change{Hydrodynamic stress fluctuations result from particle density fluctuations. The strength of the coupling is given by the inverse of the compressibility, and their time dependence results from the collective particle diffusion described in equation \eqref{eq:Strukturfactor}.}  The remainder first consists of the memory-kernel $\bf M$ which  encodes random forces and thus can be simplified in a Markovian approximation in states with weak interactions. The decomposition up to now is the expected one within the Zwanzig-Mori formalism. Yet, because of the coupling to stress fluctuations in the reducible part of the Smoluchowski operator in equation~\eqref{eq:Smolochowski_Operator_Irreducible}, a second contribution of order ${\cal O}(q^2)$ arises in the remainder. \change{It is given by the two last terms on the rhs of equation \eqref{mainResult}, and the splitting of this contribution from $M$, while not based on an exact principle, is aimed to describe slow stresses in high viscosity states. Equation~\eqref{mainResult}, which is an exact result within the projection operator formalism, thus combines fundamentally and physically motivated terms.} In fluid states, the last  two terms on the rhs in equation~\eqref{mainResult} appear negligible in the hydrodynamic limit $q\to0$ compared to the other contributions. Yet, in viscoelastic states, where the memory kernel becomes large \cite{Leutheusser,Bengtzelius}, equation~\eqref{Def_Longitudinal_Transverse_Current} shows $K_{\alpha\beta}\propto 1/q^2$, and all terms contribute comparably including in the long-wavelength limit. \change{It is noteworthy that the stress correlations in the generalized hydrodynamic limit, where all memory kernels are evaluated at $q=0$, including
\begin{align}
    M_{\alpha \beta \gamma \delta}(\pmb{0},z) &= M^{\|}(0,z) \delta_{\alpha \beta} \delta_{\gamma \delta}\\\notag
    &+M^{\perp}(0,z) \Big( \delta_{\alpha \gamma} \delta_{\beta \delta}+\delta_{\alpha \gamma} \delta_{\gamma \beta} - 2 \delta_{\alpha \beta} \delta_{\gamma \delta}\Big),
\end{align}
contain  only  two frequency dependent quantities,  the global longitudinal and shear modulus introduced in equation \eqref{eq_Longi_Trans_Memoryfunction} and being familiar from rheology \cite{Dhont}.}
\\

The expression in equation~\eqref{mainResult} is our central result and  equals the decomposition of the stress autocorrelation found in Newtonian and Langevin systems \change{with neglected hydrodynamic interactions} \cite{Vogel,Maieretall,Maieretall2}. There, the appearance of  $K_{\alpha \beta}$ arose from the coupling of the stress to the time derivative of the particle displacement, viz.~the velocity. 
It is a conserved field in Newtonian fluids, and
was included in the set of slow variables in the Langevin-case as well. The reason for this was, that the correlations of  displacements should be long-ranged and long-lived in the solid phase and close to the point of solidification. This holds for systems  immersed in a solvent as well.  While the current is no independent quantity in the overdamped Smoluchowski-dynamics, this coupling here is recovered from the proper irreducible dynamics where stress fluctuations are projected out. Again, the reasoning is that these fluctuations become slow in a viscoelastic state close to solidification. In Refs.~\cite{Vogel,Maieretall,Maieretall2}, this was modeled by a single-relaxation time approximation for the memory kernel $\bf M$, which 
introduced a description of spatial structures into Maxwell's model of a viscoelastic fluid \cite{Maxwell}. As most important result, it recovered the long-ranged stress correlations  in solid states described \change{within linearized elasticity theory} by Eshelby \cite{Eshelby1957}. As Refs.~\cite{Vogel,Maieretall,Maieretall2,Klochko2018} contain the pertinent results including the overdamped limit which is considered here, this discussion shall not be repeated. \\

\change{Note, that including  hydrodynamic interactions would give rise to additional terms decaying with time as it was worked out in \cite{Vogel}. Also, hydrodynamic long time tails and other relaxational processes will show up in the memory kernels and will differ depending on the damping of the microscopic motion. Since we are interested in the arising static properties, this discussion shall not be given here.  } 

\vspace{-0.4cm}
\section{Particle flux and viscosity}
It has been argued, that the particle displacement has to be reintroduced in the overdamped dynamics, even though the description in the Smolochowski dynamics is independent of any momenta. This \change{raise}s the question on how to define the particle flux in such systems. First, we consider the current as a linear response quantity. Via this approach, we are able relate the found memory function to the viscosity, following  \cite{HessKlein}. Secondly, we   coarse grain the Brownian motion directly and define the current from the Wiener respectively Brownian equations of motion. Both approaches yield the same result.
\vspace{-0.4cm}
\subsection{Linear response formalism \label{sey:Response_function}}
Applying a small external velocity field $\pmb{v}^{ext}(\pmb{r},t)$ gives an additional term in the  Smoluchowski equation \eqref{Smolochowski-eq} \cite{Dhont,HessKlein}
\begin{align}
\begin{split}
    \delta \Omega &= - \sum_{j=1}^N \frac{\partial}{\partial \pmb{r}_j} \cdot \pmb{v}^{ext}(\pmb{r}_j,t)\\&=- \frac{1}{(2\pi)^3}\sum_{j=1}^N \frac{\partial}{\partial \pmb{r}_j} \cdot \int d \pmb{q}'e^{i \pmb{q}'\cdot\pmb{r}_j}\, \pmb{v}^{ext} (\pmb{q}',t)\,,
\end{split}
\end{align}
 where the derivatives act on the distribution function.
 $\pmb{v}^{ext} (\pmb{q}',t)$ is the Fourier mode of the perturbation, which is essentially the Stokes' friction force with opposite sign.  The linear response theory \cite{Risken} gives for the expectation value of an arbitrary scalar dynamic variable 
\begin{align}\label{eq:Linear_Response}
\begin{split}
    \langle A(\pmb{q},t) \rangle^{\text{lr}}= -\frac{\beta}{ V} \int_{-\infty}^t d t' \braket{ A(\pmb{q}) e^{\Omega(t-t')}  \sigma^*_{\alpha \beta}(\pmb{q}) }i q_\alpha v_\beta^{ext}(\pmb{q},t) \;,
\end{split}
\end{align}
where $\braket{\cdot \cdot \cdot}^{\text{lr}}$ denotes the average over a time dependent distribution function in a linear approximation, and averages on the rhs are performed in the equilibrium, unperturbed system. This result is a manifestation of the fluctuation dissipation theorem. Translational invariance dictates that $\pmb{q}'=\pmb{q}$ holds. The Green's function $\phi_{\alpha \beta}(t-t')=\frac{\beta}{V} \braket{ A(\pmb{q}) e^{\Omega(t-t')}  \sigma^*_{\alpha \beta}(\pmb{q})} \Theta(t-t')$  is an \textit{after-effect} function, giving the response of $A$ at time $t$ to the gradient of the velocity field at time $t'$. Note that \eqref{eq:Linear_Response} can easily be generalized to non scalar quantities. Equation \eqref{eq:Linear_Response} gives the \textit{Kubo-}relation \cite{Zwanzig,nagele}
\begin{align} \label{eq:Kubo_Relation}
        \langle \sigma_{\alpha \beta}(\pmb{q}, t) \rangle^{\text{lr}}
 =-\int_{-\infty}^t d t'  C_{\alpha \beta \eta \gamma} (\pmb{q},t-t') iq_\eta  v_\gamma^{ext}(\pmb{q},t')\;. 
\end{align}
Martin \cite{Martin} or respectively  Kadanoff and Martin \cite{KardanoffMartin} suggested that  the system can still be described exclusively  by system intrinsic or local variables for small perturbation.  This suggests to  express the response function, being a functional derivative  of the responding quantity with respect to the \change{gradient} of the external velocity field, in terms of functional derivatives with respect to internal fields. 
The goal here is to identify the memory kernel as  the response of $Q \pmb{\sigma}(\pmb{q})$ to a system inherent variable $\pmb{f}$. We will  accomplish that and argue that $\pmb{f}$ can be interpreted as the \change{gradient} of the local current, meaning $f_{\alpha \beta}= \nabla_{\alpha} j_{ \beta}$. Then, following Martin and Kadanoff, the Markovian limit of the memory kernel can be identified with the transport coefficients of the hydrodynamic description.\\

Equation \eqref{eq:Linear_Response} motivates the definition of the response function of a dynamic variable $\pmb{A}$ to the \change{gradient} of $\pmb{v}^{ext}$ via the functional derivative
\begin{align}\label{def:Response-function} \begin{split} \frac{\delta \braket{\pmb{A}(\pmb{q},t)}^{\text{lr}} }{\delta i \pmb{q} \pmb{v}^{ext}(\pmb{q},t')}=- \frac{\beta}{V} \braket{\pmb{A} (\pmb{q}) e^{\Omega(t-t')} \pmb{\sigma}^*(\pmb{q})} \Theta(t-t') \;.
    \end{split}
\end{align}
The external perturbation can always be considered as a superposition of monochromatic plane waves, which factorizes in the linear response \cite{HansenMcDonald}. It is therefore sufficient to consider a single plane wave. This motivates 
a Fourier-transformation \change{ $\mathcal{F}[g(t)](\omega)=\int_{-\infty}^\infty dt e^{- i \omega t} g(t)$}, leading to
   \begin{align}
    \frac{\partial \braket{\pmb{A}(\pmb{q},\omega )}^{\text{lr}} }{i  \partial   \pmb{q} \pmb{v}^{ext}(\pmb{q},\omega)})= \frac{\beta}{V}     \braket{\pmb{A} (\pmb{q}) R(z=-i \omega) \pmb{\sigma}^*(\pmb{q})}\;,
    \end{align}
    where Cauchy's integral theorem was used. 
    Note the partial instead of the functional derivative in the frequency domain. \\
  
\change{
Motivated by  \cite{ChichockiHess, HessKlein}, we want to identify the memory kernel with the  frequency and wavevector dependent $\pmb{\eta}(\pmb{q},z=-i\omega)$  viscosity, which  is defined as the response of the out-of-equilibrium stress to the local current. But, the local current $\pmb{j}$ has to be defined as a linear response quantity, due to the overdamped description. We take 
\begin{align}
    \label{Current2}
    j_\vartheta(\pmb{q},t)=v^{ext}_\vartheta (\pmb{q},t)-i q_\tau D_0 \frac{\beta}{n} \braket{\sigma_{\tau \vartheta} (\pmb{q},t)}^{\text{lr}}
\end{align}
as a candidate.
This ansatz  translates into assuming that the local current is given by the external velocity field screened by the stress which is built up by the same perturbation.
  The agenda now is to show
  \begin{align}\label{visc_def}
    \eta_{\alpha \beta \gamma \delta}(\pmb{q},-i\omega)= \frac{\partial \braket{ Q \sigma_{\alpha \beta}(\pmb{q},-i\omega)}}{i  \partial   q_\gamma j_\delta(\pmb{q},-i\omega)}= 
        M_{\alpha \beta \gamma \delta}(\pmb{q},-i\omega)
    \;,
\end{align}
meaning that the memory function can be regarded as a generalized \textit{Green-Kubo} transport coefficient which, in accordance with \cite{HessKlein,ChichockiHess}, equals the viscosity tensor. This would support our claim, that $\pmb{M}$ is indeed the \textit{correct} Memory-function. }
\\

\change{
 The interpretation of $ j_\vartheta$ representing the local current  is based on the fact, that the divergence of  \eqref{Current2}  fulfils the continuity equation in the linear response}
 \change{
\begin{align}
     i\pmb{q} \cdot \pmb{j}(\pmb{q},t) =  \frac{1}{n} \partial_t \braket{\varrho(\pmb{q},t)}^{\text{lr}}.
\end{align}}
\change{
 To set up the continuity equation, one has to calculate the time derivative of the expectation value of the local density $\varrho(\pmb{q},t)$ in the linear response 
\begin{align}
\label{eq:Derivation_of_Density}
\partial_t \braket{\varrho(\pmb{q},t)}^{\text{lr}}=  \braket{\varrho(\pmb{q},t)\Omega}^{\text{lr}}+ \braket{\varrho(\pmb{q})\delta \Omega(t)}^{eq}.
\end{align}
Here $\delta \Omega$ represents again the perturbed Smoluchowski operator. The super-script \textit{eq} (written only in this section)  shall indicate that the average is calculated using the equilibrium distribution as denoted in equation \eqref{def_Average}. This follows from the decomposition
 $\partial_t \rho(\Gamma,t)= \Omega \rho(\Gamma,t) + \delta \Omega \rho_{eq}(\Gamma) $ which is valid in the linear approximation. One gets for the second term 
\begin{align}
    \braket{\varrho(\pmb{q})\delta \Omega}^{eq}= -i n \pmb{q} \cdot \pmb{v}^{ext}(\pmb{q},t)\;.
\end{align}
For the first term, one finds
\begin{align}
    \braket{\varrho(\pmb{q},t)\Omega}^{\text{lr}}&=   -q_{\alpha} q_\beta D_0 \beta\braket{ \sigma_{\alpha \beta}(\pmb{q},t)}^{\text{lr}}\;,
\end{align}
showing that \eqref{Current2} can indeed be interpreted as the local current. In order to relate the memory function and the response of $Q\pmb{\sigma}(\pmb{q})$ to the local current, we first analyse its response to the gradient of the external field. Using the operator identities \eqref{eq:OperatorIdentity} and \eqref{eq:Smolochowski_Operator_Irreducible}, one finds 
        \begin{align}
\label{eq:Viskosity}
\begin{split}
& \frac{\partial \braket{Q \sigma_{\alpha \beta}}^{\text{lr}}}{ i  \partial   q_\gamma v^{ext}_{ \delta}} =
   \frac{\beta}{V} \braket{\sigma_{\alpha \beta} Q R(z=- i \omega) \sigma^*_{\gamma \delta}} 
   \\&=   M_{\alpha \beta \phi \vartheta}\bigg(\delta_{\phi \gamma}\delta_{\vartheta \delta} -\frac{D_0 \beta}{n} q_\phi q_\tau \frac{\beta}{V} C_{\vartheta \tau \gamma \delta} \bigg)\;.
  \end{split}
\end{align}
Looking at \eqref{Current2}, one sees that the term in the bracket equals 
$\frac{i  \partial   q_\phi j_ \vartheta}{i  \partial   q_\gamma v^{ext}_{ \delta}}$.
Exploiting the chain rule, \eqref{eq:Viskosity} becomes}
\change{
\begin{align}
\label{eq:Viskosity2}
    \bigg(  M_{\alpha \beta \phi \vartheta} - \frac{\partial \braket{Q \sigma_{\alpha \beta}}^{\text{lr}}}{i  \partial   q_\phi j_ \vartheta} \bigg) \frac{i  \partial   q_\phi j_ \vartheta}{i  \partial   q_\gamma v^{ext}_{ \delta}}= \left.\frac{\partial \braket{Q \sigma_{\alpha \beta}}^{\text{lr}}}{ i  \partial   q_\gamma v^{ext}_{ \delta}} \right|_{i \pmb{q} \cdot \pmb{j}=const}.
\end{align}}
\change{
Up to this point, our argumentation was basically, that we need to reintroduce the local current in our set of distinguished quantities, even though it is not a dynamical variable in the present framework. Relying on Martin's and Kardanoff's suggestion  once more, and keeping in mind that the density is the only other distinguished variable in our model, we postulate that
 the right hand side of \eqref{eq:Viskosity2} vanishes. An external velocity field causes an internal particle current, which then builds up stresses. The vanishing of 
the right hand side of \eqref{eq:Viskosity2}
then requires the bracket on the left hand side to vanish as well.} \\

\change{
So the memory kernel can indeed be interpreted as the response function of the projected stress tensor to the local current. 
Meaning that the memory function can be identified with a generalized \textit{Green-Kubo} transport coefficient, which is the viscosity tensor in the present case \eqref{visc_def}.  In the limit of long wavelengths and small frequencies, it approaches the viscosity  as the irreducible dynamics simplifies, viz.~$R^{irr}(z) \to Q R(z) Q$ for $q\to0$; this follows from equation~\eqref{mainResult}. The expected Green-Kubo relation holds \cite{HansenMcDonald}. }

\subsection{Hydrodynamic equations}
The hydrodynamic description of the slow dynamics of a Brownian system shall be obtained by coarse-graining the equations of motion, being  the  set of  overdamped Langevin, respectively Brownian or Wiener  equations  \cite{Dhont}
\begin{align}
    \label{Langevin}
    \gamma_0\left(\dot{\pmb{r}}_j(t)-\pmb{v}^{\text{ext}}(\pmb{r}_j,t)\right) =\pmb{F}_j(\Gamma)+\pmb{f}_j(t)\;,
\end{align}
where the random noise $\pmb{f}_j(t)$ is Gaussian and white, and obeys $\braket{\pmb{f}_j(t),\pmb{f}_k(t')}=k_B T\gamma_0 \delta(t-t')\delta_{jk}$.  Here, $ \pmb{F}_j(\Gamma) $ is the total force acting on the $j^{th}$ particle caused by the interaction with the remaining colloids. Equation~\eqref{Langevin} describes particles performing random walks relative to a flowing background. The difference $\dot{\pmb{r}}_j(t)-\pmb{v}^{\text{ext}}(\pmb{r}_j,t)$ gives the non-affine motion. Using the Kramers-Moyal expansion \cite{Risken} one verifies that the evolution of the system is equivalent to the one described by  the Smoluchowski-equation \eqref{Smolochowski-eq}. Using a coarse-graining approach \cite{Goldhirsch2010} we define the density field as 
\begin{align}
    \label{CGDensity}
    \varrho(\pmb{r},t)=\sum_{j=1}^N \phi(\pmb{r}-\pmb{r}_j(t))
\end{align}
and the particle flux as
\begin{align}
    \label{CGflux}
    \pmb{j}(\pmb{r},t)=\sum_{j=1}^N \dot{\pmb{r}}_j(t) \phi(\pmb{r}-\pmb{r}_j(t))\;.
\end{align}
The coarse-graining function $\phi(\pmb{r})$ can be \change{pictured as being} Gaussian with the width $w$ and normalization $\int d\pmb{r} \phi(\pmb{r})=1.$ \change{A smooth $\phi(\pmb{r})$ is considered in order to prepare the application of the approach to  simulations, while a Dirac delta leads back to the field definitions in the earlier sections such as equation \eqref{eq:def_density}.   }
Note that the density in equation~\eqref{CGDensity} obeys the continuity equation, \change{ $\partial_t  \varrho+\nabla\cdot \pmb{j}=0$}. Inserting the Brownian equation of motion \eqref{Langevin} into the definition of the flux \eqref{CGflux} and using Newton's third law gives the coarse grained stress tensor
\begin{align}
    \label{CGStress}
    \begin{split}
    -\nabla \cdot \pmb{\sigma}&(\pmb{r},t)\coloneqq\gamma_0\Big(\pmb{j}(\pmb{r},t)-n\pmb{v}^{\text{ext}}(\pmb{r},t)\Big)- \mathcal{F}(\pmb{r},t)\\
    &=\frac{1}{2}\sum_{j\neq k} \pmb{F}_{jk}\Big(\phi(\pmb{r}-\pmb{r}_j(t))-\phi(\pmb{r}-\pmb{r}_k(t)\Big)\;.
        \end{split}
\end{align}
The fluctuation force $\mathcal{F}=\sum \pmb{f}_j(t)\phi(\pmb{r}-\pmb{r}_j(t))$ will be neglected in the following. Note, that the equivalence in \eqref{CGStress} is not exact. We rather used a \change{saddlepoint approximation} for the external velocity field. Correction terms will arise if the external velocity varies rapidly on the scale of the particle interactions. Equation \eqref{CGStress} defines the coarse grained stress tensor up to an addable, divergence free term
\begin{align}
\label{def:CGStress}
\begin{split}
\pmb{\sigma}(\pmb{r},t) = \frac{1}{2} \sum_{k\neq j}  \pmb{r}_{jk}(t)  \pmb{F}_{jk}(\Gamma)\; \int_0^1\!\!\!\!ds  \phi\big(\pmb{r}-\pmb{r}_j(t)+s \pmb{r}_{jk}(t)\big)\;.  
\end{split}
\end{align}
The diagonal elements of the stress tensor are used to define the local pressure $p$. This motivates the decomposition
\begin{align}
    \label{DecomCGStress}
    \pmb{\sigma}(\pmb{r},t)=p(\pmb{r},t) \mathds{1}- \delta \Tilde{\pmb{\sigma}}(\pmb{r},t)\;.
\end{align}
Where the deviatoric stress tensor $\delta \Tilde{\pmb{\sigma}}(\pmb{r},t)$ is caused by viscous forces. The pressure varies with the local density \cite{HessKlein,HansenMcDonald} according to $p(\pmb{r},t)=p_{eq}+ \frac{k_B T}{S_0} \delta \varrho(\pmb{r},t)$, \change{ where $nk_BT/S_0$ is the inverse isothermal compressibility}.
Assuming local thermodynamic equilibrium, the off-diagonal elements of the stress tensor are related to a perturbing external velocity field in the hydrodynamic limit of small wavevectors. This motivates the following identification to connect the hydrodynamic description to the one based on correlation-functions (Sect. IV): 
\begin{align}
    \label{IndetificationCGSTrresLRQStress}
    \delta \Tilde{\pmb{\sigma}}(\pmb{r},t)=\braket{Q\pmb{\sigma}(\pmb{q}\to 0,t}^{\text{lr}}_{(\pmb{r},t)}.
\end{align}
The notation $. . . (\pmb{q} \to 0, t)\rangle^{\text{lr}}_{(\pmb{r},t)} $ implies that the coarse-graining size $w$ is so large that in the evaluation of linear response functions all particle correlations have been integrated and that a spatial variation only remains because of the slow variation of the external fields. \change{Note that the signs in \eqref{DecomCGStress} and \eqref{IndetificationCGSTrresLRQStress} are motivated by the linear response consideration \eqref{eq:Linear_Response}.} 
\\

Following \cite{HessKlein}, we define the viscosity as a generalized transport coefficient for the stress fluctuations
\begin{align}
\label{Def_Viscosity_Hydrodynamic}
 \delta \Tilde{{\sigma}}_{\alpha \beta}(\pmb{r},t)=\int_{-\infty}^t dt' \eta_{\alpha \beta \gamma \delta}(t-t')\nabla_\gamma j_\delta(\pmb{r},t)\;.
\end{align}
As suggested by Martin and Kadanoff \cite{Martin,KardanoffMartin}, this constitutive equation defines the viscosity via the response of the stress to the gradient of the internal current field.
It is more convenient to express the response function again in the frequency domain 
\begin{align}
    \label{FreWaveViscosity}
    n \frac{\delta \,( \delta \Tilde{{\sigma}}_{\alpha \beta}(\pmb{r},z))}{\delta \nabla_\gamma j_\delta(\pmb{r}',z)} = \eta_{\alpha \beta \gamma \delta}(z)\delta(\pmb{r}-\pmb{r}')\;.
\end{align}
Note, that the viscosity is defined in the limit $\pmb{q}\to 0.$
With this and equation \eqref{CGStress} we regain the constitutive equation for the viscosity \eqref{visc_def} from the projection formalism approach via the present hydrodynamic framework (the fluctuation force is neglected). Equation \eqref{eq:Viskosity} now reads
\begin{widetext}
\begin{align}
\label{Viscosity2}
\begin{split}
    \frac{\delta( \delta \Tilde{{\sigma}}_{\alpha \beta}(\pmb{r},z))}{\delta \nabla_\gamma v^{ex}_\delta(\pmb{r}',z)}&= \frac{\delta (\delta \Tilde{{\sigma}}_{\alpha \beta}(\pmb{r},z))}{\delta \nabla_\xi j_\tau(\pmb{r}'',z)} \frac{\delta \nabla_\xi j_\tau(\pmb{r}'',z)}{\delta \nabla_\gamma v^{ex}_\delta(\pmb{r}',z)} \\
      & = \eta_{\alpha \beta \xi \tau }(z)\delta(\pmb{r}-\pmb{r}'')\bigg[\delta_{\xi \gamma }\delta_{\tau \delta}\delta(\pmb{r}-\pmb{r}')-\frac{D_0}{S_q n} \nabla_\xi\frac{\delta \nabla_\tau p(\pmb{r}'',z)}{\delta \nabla_\gamma v^{ex}_\delta(\pmb{r}',z)} - \nabla_\xi \frac{\delta \nabla_\zeta \delta \Tilde{{\sigma}}_{\zeta \tau}(\pmb{r}'',z)}{\delta \nabla_\gamma v^{ex}_\delta(\pmb{r}',z)} \bigg]\;.
\end{split}
\end{align}
\end{widetext}
The equations \eqref{CGStress} and \eqref{Viscosity2} \change{(in the Markovian limit)} lead to the final hydrodynamic equation for the particle current 
\begin{align}
    \label{HydroFluc}
\pmb{j}(\pmb{r},t) - \frac{1}{n\gamma_0}\; \nabla \left( \pmb{\eta} : \nabla \; \pmb{j}(\pmb{r},t) \right)  = n \pmb{v}^{ ext}(\pmb{r},t) - \frac{1}{\gamma_0}\, \nabla p(\pmb{r},t).
\end{align}
This is equivalent to equation \eqref{Current2}. So, the hydrodynamic approach gives the same result as the \textit{Zwanzig-Mori} projection formalism. This supports the claim that the memory function \eqref{eq:Memoryfunction} is indeed the correct one, since it can be interpreted as the viscosity in both approaches. 

The hydrodynamic equation \eqref{HydroFluc}
generalizes the one for an incompressible fluid given in \cite{Vogel}, which was recently tested in simulations of the Stokes-friction \cite{Orts2020}. Together with the conservation law of the density, it captures the linearized \change{generalized} hydrodynamic regime of a fluid of interacting Brownian particles. \change{In the true hydrodynamic limit, density diffusion results from the leading gradient,  $\pmb{j}(\pmb{r},t)- n \pmb{v}^{ ext}(\pmb{r},t)\to - (k_BT/\gamma_0S_0) \nabla \varrho(\pmb{r},t)$.} In \cite{Maieretall2,Maieretall} following the strategy going back to  Maxwell, the \change{approximation of} generalized hydrodynamics capturing viscoelastic Newtonian fluids  was discussed. This generalized hydrodynamics can easily be transferred to equation~\eqref{HydroFluc} assuming a frequency dependence of the shear and longitudinal viscosities in $\pmb{\eta} $. \change{(This is equivalent to keeping the convolution in equation \eqref{Def_Viscosity_Hydrodynamic}.) In the solid limit, where the velocity field is the time derivative of a displacement field, $\pmb{j}(\pmb{r},t)=n\dot{\pmb{u}}(\pmb{r},t)$, this leads to the linearized static equations of elasticity theory $\nabla \left( \delta \tilde{\pmb{\sigma}}(\pmb{r},t)-p(\pmb{r},t)\mathds{1} \right)=-\gamma_0n\pmb{v}^{\rm ext}$, with the Hookean stress of an isotropic solid,  $\delta\tilde{\sigma}_{\alpha\beta}= (M^\|_\infty-2M^\perp_\infty) (\nabla\cdot \pmb{u}) \delta_{\alpha\beta}+M^\perp_\infty(\nabla_\alpha u_\beta+\nabla_\beta u_\alpha)$ and  the rhs as an external source of forcing \cite{Picard2004}. Here, $M^\|_\infty$ and $M^\perp_\infty$ are the elastic contributions in the longitudinal and shear modulus. It is the potential to bridge between both limits, the hydrodynamic fluid and the Hookean solid one, which we consider the strength of the presented generalized hydrodynamics.}

%
\section{Conclusions}
Employing the projection operator formalism, we decomposed the stress autocorrelation in Brownian systems into a structure that formally agrees with the one previously obtained in Newtonian  \cite{Maieretall2,Maieretall} or  Langevin systems \cite{Vogel}. In those systems the dynamical coupling between stresses and momentum currents was considered, while particle momenta are not among the dynamical variables in the Brownian case. This interpretation is based on the fact, that in the final expression for the correlation function \eqref{mainResult} a matrix $K_{\alpha \beta}$ appears which is identical to the autocorelation of the current in the overdamped Langevin-system. So, as one would expect, it makes no difference whether the calculation is done in a general Langevin-system with the Fokker-Planck operator and the overdamped approximation is made at the end, or whether one directly starts in the overdamped Smoluchowski system. In both cases, the same coupling of the stress to the current, or respectively to the time derivative of the displacement field, occurs. It has to be included in a generalized hydrodynamics which aims to capture viscoelastic states and the solid limit \cite{Maieretall2}. {  Starting on the level of the Smoluchowski equation elaborates the role of stress correlations, which manifestly enter the definition of the irreducible dynamics.} \\

Furthermore, we generalized the consideration by Cichoki and Hess \cite{ChichockiHess} and Kawasaki \cite{Kawasaki} for the memory function of the dynamic structure factor by including transverse contributions as well.  With this, we were able to generalize their linear response argument. The obtained memory function gives the response of the stress to the internal particle current and thus, following Kadanoff and Martin \cite{Martin,KardanoffMartin}, can be interpreted as the generalized viscosity tensor. Additionally, this result was obtained via a hydrodynamic approach. The final hydrodynamic equation for the particle current is consistent with the one obtained in the linear response formalism. 

As additional result, we obtained the hydrodynamic equation for the particle current 
in a Brownian fluid. The equation can be considered the analogue of the Navier-Stokes equation for a Newtonian fluid.  \change{Determining the particle current $\pmb{j}(\pmb{r},t)$ is also the aim of dynamic density functional theory for Brownian systems (DDFT) \cite{Marconi1999}. Its expression reads $\gamma_0 \pmb{j}(\pmb{r},t) = -  \varrho(\pmb{r},t) \nabla \frac{\delta {\cal F}}{\delta  \varrho(\pmb{r},t)} $ where $\cal F$ is the free energy functional.  Differently from the coarse grained equation \eqref{HydroFluc}, the density field in DDFT is an ensemble averaged quantity that varies on microscopic length scales. Power functional theory \cite{Schmidt2013} is a generalization of DDFT which appears closer in structure to equation \eqref{HydroFluc} especially in the velocity gradient formalism \cite{delasHeras2018}, and should be compared in the long wavelength limit.} 

\section{Acknowledgements}

Open Access Funding enabled and organized by Projekt
DEAL.
We are indebted to Prof. Dr. Annette Zippelius for helpful discussions  and a careful reading of the manuscript, and acknowledge financial support from the German Science Foundation (DFG) under project no.  FU 309/12.

\section{Author contribution statement}

Both authors were involved in the research and in
the preparation of the manuscript. Both authors have read and approved the
final manuscript.

\medskip

\bibliography{sources}

\end{document}